\documentclass[preprints, discussion, accept, oneauthor, pdftex]{mdpi}
\usepackage{lmodern}
\usepackage{amssymb,amsmath}
\usepackage{ifxetex,ifluatex}
\usepackage{fixltx2e} 
\ifnum 0\ifxetex 1\fi\ifluatex 1\fi=0 
  \usepackage[T1]{fontenc}
  \usepackage[utf8]{inputenc}
\else 
  \ifxetex
    \usepackage{mathspec}
  \else
    \usepackage{fontspec}
  \fi
  \defaultfontfeatures{Ligatures=TeX,Scale=MatchLowercase}
\fi
\IfFileExists{upquote.sty}{\usepackage{upquote}}{}
\IfFileExists{microtype.sty}{%
\usepackage{microtype}
\UseMicrotypeSet[protrusion]{basicmath} 
}{}
\usepackage{hyperref}
\hypersetup{unicode=true,
            pdftitle={Pearle's Hidden-Variable Model Revisited},
            pdfauthor={Richard D. Gill (Leiden University)},
            pdfborder={0 0 0},
            breaklinks=true}
\usepackage{color}
\usepackage{fancyvrb}

\DefineVerbatimEnvironment{Highlighting}{Verbatim}{commandchars=\\\{\}}
\usepackage{framed}
\definecolor{shadecolor}{RGB}{248,248,248}
\newenvironment{Shaded}{\begin{snugshade}}{\end{snugshade}}

\newcommand{\CommentTok}[1]{\textcolor[rgb]{0.56,0.35,0.01}{\textit{#1}}}

\newcommand{\ControlFlowTok}[1]{\textcolor[rgb]{0.13,0.29,0.53}{\textbf{#1}}}
\newcommand{\DataTypeTok}[1]{\textcolor[rgb]{0.13,0.29,0.53}{#1}}
\newcommand{\DecValTok}[1]{\textcolor[rgb]{0.00,0.00,0.81}{#1}}

\newcommand{\FloatTok}[1]{\textcolor[rgb]{0.00,0.00,0.81}{#1}}

\newcommand{\KeywordTok}[1]{\textcolor[rgb]{0.13,0.29,0.53}{\textbf{#1}}}
\newcommand{\NormalTok}[1]{#1}
\newcommand{\OperatorTok}[1]{\textcolor[rgb]{0.81,0.36,0.00}{\textbf{#1}}}

\newcommand{\StringTok}[1]{\textcolor[rgb]{0.31,0.60,0.02}{#1}}

\usepackage{graphicx,grffile}
\makeatletter
\def\maxwidth{\ifdim\Gin@nat@width>\linewidth\linewidth\else\Gin@nat@width\fi}
\def\maxheight{\ifdim\Gin@nat@height>\textheight\textheight\else\Gin@nat@height\fi}
\makeatother
\setkeys{Gin}{width=\maxwidth,height=\maxheight,keepaspectratio}
\ifxetex
  \usepackage[setpagesize=false, 
              unicode=false, 
              xetex]{hyperref}
\else
  \usepackage{hyperref}
\fi

\firstpage{1} 
\pubvolume{xx}
\issuenum{1}
\articlenumber{5}
\pubyear{2019}
\copyrightyear{2019}
\history{Received: date; Accepted: date; Published: date}
\pdfoutput=1

\def\linenumbers{{}}    

\Title{Pearle's Hidden-Variable Model Revisited}

\Author{Richard David Gill $^{1}$\orcidA{}}

\AuthorNames{Richard Gill}

\address[1]{%
$^{1}$ \quad Leiden University, Faculty of Science, Mathematical Institute; gill@math.leidenuniv.nl}

\corres{Correspondence: e-mail@e-mail.com; Tel.: (optional; include country code; if there are multiple corresponding authors, add author initials) +xx-xxxx-xxx-xxxx (F.L.)}

\abstract{Pearle (1970) gave an example of a local hidden variables model which exactly reproduced the singlet correlations of quantum theory, through the device of data-rejection: particles can fail to be detected in a way which depends on the hidden variables carried by the particles and on the measurement settings. If the experimenter computes correlations between measurement outcomes of particle pairs for which both particles are detected, he or she is actually looking at a subsample of particle pairs, determined by interaction involving both measurement settings and the hidden variables carried in the particles. We correct a mistake in Pearle's formulas (a normalization error) 
and more importantly show that the model is more simple than first appears. 
We illustrate with visualisations of the model and with a small simulation experiment, with code in the statistical programming language R included
in the paper. Open problems are discussed.}


\begin{document}

\maketitle

\noindent {{\textbf{Preliminary note}: this paper has been composed in the {\em R markdown language} and typeset using the {\em R} package ``{\tt knitr}'' and {\em RStudio}, a popular IDE for working with {\em R}. The source code contains therefore interleaved passages of text (including {\LaTeX} code, for instance, for mathematical formulas) and {\em R} code. Processing the original {\em Rmd} file with {\tt knitr} generates a {\LaTeX} source file containing interleaved text, {\em R} code and {\em R} textual output. It also generates pdf figures of {\em R}'s graphical output. Some minor editing of the {\LaTeX} preamble was necessary to typeset in MDPI's house style. Finally {\tt pdflatex} generates the pdf you are reading now. The source file is available from the author.

Most readers will not be interested in the R code. It can easily be skipped; hopefully the figures are well enough explained in the surrounding text.}

\section{Introduction}

Bell's (1964) landmark paper ``On the Einstein Podolsky Rosen paradox'' led a few years later to a version of his inequality more suitable for experimental purposes, and consequently the focus of a very great deal of both experimental and theoretical work. 
That is the inequality nowadays called the Bell-CHSH inequality, presented by Clauser, Horne, Shimony and Holt (1969). Almost immediately, however, Pearle (1970) pointed out that the problem of detector efficiency meant that it was easy under local realism to reproduce the famous negative cosine curve of the correlations between spin measurements on particles in the singlet state. The measurements on each particle would not have two outcomes but three: spin up, spin down, and no detection. One would be tempted to restrict attention to only those ``trials'' in which both particles were detected, and compute the correlation between the observed spins for that subpopulation. Whether or not a particle was detected was, in Pearle's
model, determined by a hidden variable correlated with the actual ``hidden'' spins of the particles. Detection depended on the extra hidden variable and on the detector setting. Selection of particle pairs such that both particles got detected effectively selects a subpopulation of particle pairs, whose hidden spins actually depend on the detector settings.

This would result in experimental violation of the CHSH inequality, moreover with the maximal violation predicted by quantum mechanics, even though there is a perfect local realistic explanation of the correlations found.

Pearle's model is the subject of this paper. It was the starting shot in a huge literature on the detection loophole, which continues to grow to this day. Pearle's model did have some unphysical features, and he was well aware of them. In his model, the probability of a double detection would depend on the angle between the two detectors and hence the experimenter would immediately notice that his or her results did not make sense. The paper was for many years considered a purely theoretical exercise which established a purely theoretical lower limit to detector efficiency which would have to be exceeded before a so-called loophole-free experiment could be carried out. Soon, other detection loophole models were discovered which did not have his model's defect. Later, such models were found which moreover established the same lower limit, and it was also shown that the bound was optimal.

Since the literature is so huge, it cannot be adequately surveyed in this paper, and I refer the reader to the most recent comprehensive survey, Larsson (2014). That paper in fact covers all of the ``known'' loopholes, not just the detection loophole. A year later, in 2015, the first ``loophole-free'' experiments were performed and experimental violation of appropriate Bell-type inequalities observed. Yet the detection loophole remains of great interest and new detection-loophole models are continually being invented. In fact, many, both old and new, can even
be considered as variations on Pearle's. The purpose of this paper is to clarify this situation, and also to make Pearle's work more accessible. His paper is unfortunately marred by curious notational conventions, confusing misprints, and some real errors in key formulas (incorrect normalization constants). It seems that these errors have not been noticed before. In fact, as far as I know, nobody had actually tried to implement Pearle's model in simulation programs before. (Philip Pearle himself, private communication, was also not aware of any implementation).

I will just mention a very small number of other key papers. Regarding the early years, 
the landmark paper Clauser \& Horne (1974) already includes another detection-loophole model
without the just mentioned bad feature of Pearle's. Later, an important survey and many new results were
provided by Garg \& Mermin (1987). A little known but very interesting survey was provided by Risco-Delgado (1993).
A whole series of important contributions was made by Jan–\AA ke Larsson (see his survey paper); 
of particular relevance to the detector efficiency issue 
are Larsson (1998) and Larsson \& Semitecolos (2001).

\section{Pearle's model simplified}

Pearle's model is best understood with the help of a picture. The following is
taken (and used with the author's permission) from Risto-Delgado (1993).

\includegraphics{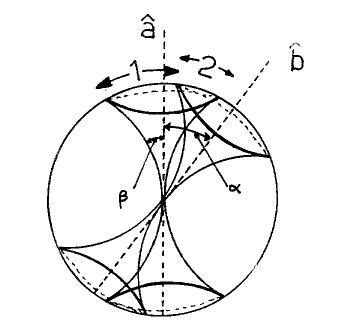}

\noindent Risto-Delgado explains as follows: Pearle’s sphere represents the nine possible outcomes of an EPR experiment allowing undetected events. If the representative point lies in region 1, the particle A will be measured as possessing spin parallel to $\hat a$, and the particle B will not be detected. If it lies in region 2 the particle the particle A will be measured as possessing spin parallel to $\hat a$ and particle B antiparallel to $\hat b$, and so on. There exist nine possibilities defending on the position of the point and the relative angle $\alpha$ between $\hat a$ and $\hat b$.

Pearle, and Risto-Delgado, are modelling a source emitting pairs of particles. The
particles carry hidden variables \(\mathbf X\) and \(\mathbf Y\) which
we take to be random points in the unit ball in \(\mathbb R^3\). We
assume that \({\mathbf Y} = - {\mathbf X}\) and
\(\mathbf X\ne \mathbf 0\) with probability 1. The ball is drawn in the figure.

Write \({\mathbf X} = R {\mathbf U}\) where \(\|{\mathbf U}\|^2 = 1\) and
\(R > 0\). One might think of the unit length vector \(\mathbf U\) as
the \emph{direction} of spin of the first particle, and \(-\mathbf U\)
as the direction of spin of the second, equal and opposite points on the
unit sphere \(S^2\), while the scalar \(R\) is some kind of amplitude of
spin.

Assume that the direction \(\mathbf U\) is uniformly distributed on
\(S^2\) and statistically independent of the amplitude \(R\in(0, 1]\).

Notation: bold (as opposed to italic) indicates a vector; random vectors
and random variables are denoted by upper case symbols, while lower case
is used for non-random quantities.

Each particle gets measured in directions \(\mathbf a\) and
\(\mathbf b\) respectively (points on \(S^2\), chosen freely by the
experimenter); these were the directions $\hat a$ and $\hat b$ in the figure.
The possible outcomes are \(+1\) (``spin up''), \(-1\)
(``spin down''), and last but not least ``no detection'', according to
the following rule: if the angle between \(\mathbf X\) and \(\mathbf a\)
is less than \(R \pi/2\) then the outcome of measuring the first
particle is \(+1\); if the angle between \(\mathbf X\) and
\(- \mathbf a\) is less than \(R\pi/2\) then the outcome of measuring
the first particle is \(-1\); otherwise the particle is not detected at
all. The rule for the second particle is exactly the same story as for
the first particle, with \(\mathbf X\) and \(\mathbf a\) replaced
throughout by \(\mathbf Y\) and \(\mathbf b\).

The smaller of the angles between \(\mathbf X\) and \(\pm\mathbf a\) is
\(\cos^{-1} \left|\mathbf U \cdot \mathbf a\right|\) so the recipe
becomes: the outcome of measuring the first particle is
\(\textrm{sign}\, \mathbf U \cdot \mathbf a\) if
\(\cos^{-1} \left|\mathbf U \cdot \mathbf a\right| \ge R\pi/2\) while
there is ``no detection'' if
\(\cos^{-1} \left|\mathbf U \cdot \mathbf a\right| < R\pi/2\); the
outcome of measuring the second particle is
\(- \textrm{sign}\, \mathbf U \cdot \mathbf b\) if
\(\cos^{-1} \left|\mathbf U \cdot \mathbf b\right| \ge R\pi/2\) while it
is not detected if
\(\cos^{-1} \left|\mathbf U \cdot \mathbf b\right| < R\pi/2\).

Pearle (1970) gives a formula, (22) in his paper, for a particular
choice of the probability density of \(R\); but take note of his
idiosyncratic normalisation (1)! There is an error in his derivation,
as can be verified by integrating the density over the whole range:
combining (1) and (22), we get a density which does not integrate to
\(1\). Working through Pearle's paper in detail, it turns out that the
only error in (22) is the normalisation constant, and this probably
derives from an incorrect normalization in (14) where Pearle switches
from \(R\) to \(S = \cos( R \pi/2)\), but it is difficult to be certain
about this, since his notion of probability density is ambiguous and
unconventional.

Here I will present an alternative and much simpler description of the
distribution of \(R\) and also of the whole model, via the distribution
of \(S = \cos( R \pi/2) \in [0, 1)\). It turns out that the distribution
of \(S\) can be expressed by the formula \(S = (2/\sqrt V) - 1\) where
\(V\sim\textrm{Unif}(1,4)\); and moreover it is \(S\) which we primarily
need to know in order to simulate the model.

In terms of \(S = (2/\sqrt V) - 1\), the recipe for simulating the
measurement of one pair of particles is as follows: generate
\(\mathbf U\) uniformly at random on the sphere \(S^2\) and
independently thereof, generate \(V\) uniformly at random in the real
interval \([1, 4]\). Compute \(S = (2/\sqrt V) - 1 \in (0, 1)\),
\(A = \mathbf U \cdot \mathbf a\), and
\(B = - \mathbf U \cdot \mathbf b\). Particle 1 is detected if and only
if \(\left | A \right | \ge S\), and if it is detected, the outcome of
measurement is \(\textrm{sign}(A)\). Particle 2 is detected if and only
if \(\left | B \right | \ge S\), and if it is detected, the outcome of
measurement is \(\textrm{sign}(B)\).

Pearle's main result is that this model reproduces the singlet
correlations:
\[\textrm{E}\Bigl(\textrm{sign}(A)\textrm{sign}(B) \Bigm| \left | A \right | \ge S ~\textrm{and}~ \left | B \right | \ge S\Bigr)~=~-\mathbf a \cdot \mathbf b.\]

I do not reproduce Pearle's (magnificent but of necessity very involved)
proof. Instead I will just derive the density of \(R\) according to my
specification, so that the reader can compare with Pearle's formula. I
will then ``prove'' Pearle's result by a simulation experiment. In fact,
I would dearly like to see a short-cut derivation of Pearle's result.
Through some quite brilliant calculations, he characterizes all possible
probability distributions of \(R\) (equivalently, of \(S\)) which will
reproduce the singlet correlations as (up to normalization) the positive
functions within the range of a certain differential operator, and then
shows that the operator when applied to the constant function -- the
most simple choice one could make -- is indeed positive. Further details
are given in an appendix at the end of this paper.

According to my definitions, \(R = \cos^{-1}(S) / (\pi/2)\) and it
follows that for \(r \in (0, 1)\),
\[\Pr(R \le r)~ =~ \Pr(S \ge \cos(r \textstyle{\frac \pi 2})) ~=~ \Pr(2/\sqrt V - 1 \ge \cos(r \textstyle{\frac \pi 2}))\]
\[=~\Pr\Biggl(\sqrt V  \le \frac 2 {1 + \cos(r \frac \pi 2)}\Biggr)\]
\[=~\Pr\Biggl( V  \le \frac 4 {(1 + \cos(r \frac \pi 2))^2}\Biggr)\]
\[=~\frac 1 3 \Biggl( \frac 4 {(1 + \cos(r \frac \pi 2))^2} - 1\biggr)\]
and hence the probability density of \(R\) is
\[f_R(r)~=~\frac 43 \cdot 2 \cdot \frac \pi 2 \cdot\frac{ \sin(r \frac \pi 2)}{(1 + \cos(r \frac \pi 2))^3} \]
\[~=~\frac {4\pi}3  \cdot\frac{ \sin(r \frac \pi 2)}{(1 + \cos(r \frac \pi 2))^3} \]
on the interval \((0, 1)\). Compare this to Pearle's formulas (1) and
(22) combined:
\[4 \pi \rho(r)r^{2}~=~ \frac{16} 3 \cdot \frac{ \sin(r \frac \pi 2)}{(1 + \cos(r \frac \pi 2))^3} .\]

Here is a graph of the probability density of \(R\), together with a
rough numerical check that it integrates to \(1\). Since the probability
density is monotone increasing we get guaranteed lower and upper bounds
to the integral by summing the value of the density on a regular grid of
points between \(0\) and \(1\), omitting the right hand and left hand
endpoints respectively, and dividing by the number of intervals
generated by the grid.

\begin{Shaded}
\begin{Highlighting}[]
\NormalTok{r <-}\StringTok{ }\KeywordTok{seq}\NormalTok{(}\DataTypeTok{from =} \DecValTok{0}\NormalTok{, }\DataTypeTok{to =} \DecValTok{1}\NormalTok{, }\DataTypeTok{length =} \DecValTok{1001}\NormalTok{)}
\NormalTok{f <-}\StringTok{ }\NormalTok{(}\DecValTok{4} \OperatorTok{*}\StringTok{ }\NormalTok{pi }\OperatorTok{/}\StringTok{ }\DecValTok{3}\NormalTok{) }\OperatorTok{*}\StringTok{ }\KeywordTok{sin}\NormalTok{(r }\OperatorTok{*}\StringTok{ }\NormalTok{pi }\OperatorTok{/}\StringTok{ }\DecValTok{2}\NormalTok{) }\OperatorTok{/}\StringTok{ }\NormalTok{(}\DecValTok{1} \OperatorTok{+}\StringTok{ }\KeywordTok{cos}\NormalTok{(r }\OperatorTok{*}\NormalTok{pi }\OperatorTok{/}\StringTok{ }\DecValTok{2}\NormalTok{))}\OperatorTok{^}\DecValTok{3}
\KeywordTok{plot}\NormalTok{(r, f, }\DataTypeTok{bty =} \StringTok{"n"}\NormalTok{, }\DataTypeTok{type =} \StringTok{"l"}\NormalTok{,}
  \DataTypeTok{main =} \KeywordTok{bquote}\NormalTok{(}\StringTok{"Probability density of"} \OperatorTok{~}\StringTok{ }\KeywordTok{italic}\NormalTok{(R)),  }
  \DataTypeTok{xlab =} \KeywordTok{bquote}\NormalTok{(}\StringTok{"Radius"} \OperatorTok{~}\StringTok{ }\KeywordTok{italic}\NormalTok{(r)),}
  \DataTypeTok{ylab =} \KeywordTok{bquote}\NormalTok{(}\StringTok{"Density"} \OperatorTok{~}\StringTok{ }\KeywordTok{italic}\NormalTok{(f)),}
  \DataTypeTok{sub =} \StringTok{"(Check: area under curve = 1)"}\NormalTok{)}
\end{Highlighting}
\end{Shaded}

\includegraphics{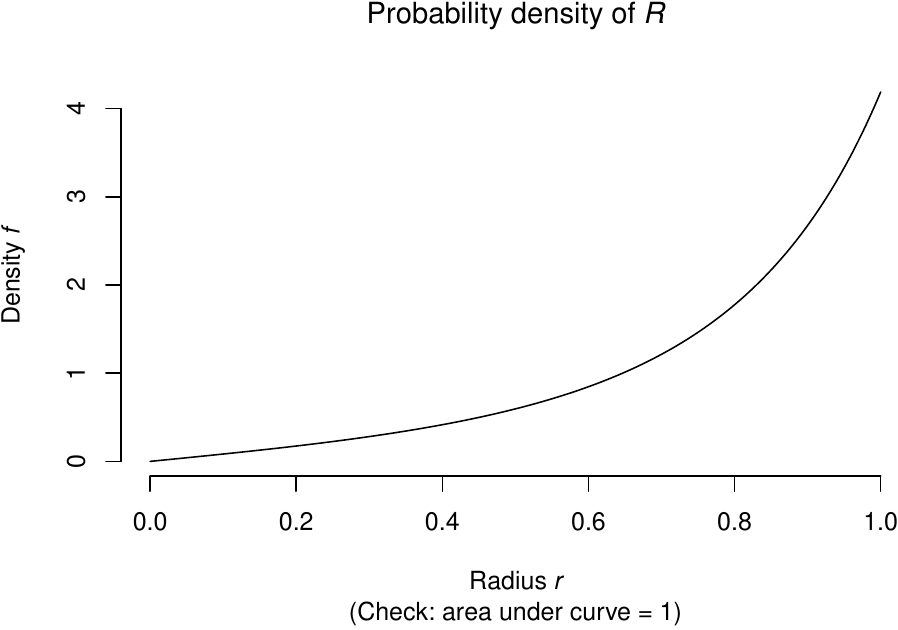}

\begin{Shaded}
\begin{Highlighting}[]
\KeywordTok{sum}\NormalTok{(f[}\DecValTok{1}\OperatorTok{:}\DecValTok{1000}\NormalTok{] }\OperatorTok{/}\StringTok{ }\DecValTok{1000}\NormalTok{)}
\end{Highlighting}
\end{Shaded}

\begin{verbatim}
## [1] 0.9979072
\end{verbatim}

\begin{Shaded}
\begin{Highlighting}[]
\KeywordTok{sum}\NormalTok{(f[}\DecValTok{2}\OperatorTok{:}\DecValTok{1001}\NormalTok{] }\OperatorTok{/}\StringTok{ }\DecValTok{1000}\NormalTok{)}
\end{Highlighting}
\end{Shaded}

\begin{verbatim}
## [1] 1.002096
\end{verbatim}

If the points \(X\) had been chosen uniformly distributed within the
unit ball, the probability density of their distance \(R\) to the orgin
would have had probability density \(3 r^2\), \(0 \le r \le 1\). In the
next plot I compare the two densities (the one corresponding to a
uniform distribution over the ball in green).

\begin{Shaded}
\begin{Highlighting}[]
\NormalTok{g <-}\StringTok{ }\DecValTok{3} \OperatorTok{*}\StringTok{ }\NormalTok{r}\OperatorTok{^}\StringTok{ }\DecValTok{2}
\KeywordTok{plot}\NormalTok{(r, f, }\DataTypeTok{bty =} \StringTok{"n"}\NormalTok{, }\DataTypeTok{type =} \StringTok{"l"}\NormalTok{,}
  \DataTypeTok{main =} \KeywordTok{bquote}\NormalTok{(}\StringTok{"Comparison of two models for density of"}\OperatorTok{~}\StringTok{ }\KeywordTok{italic}\NormalTok{(R)),  }
  \DataTypeTok{xlab =} \KeywordTok{bquote}\NormalTok{(}\StringTok{"Radius"} \OperatorTok{~}\StringTok{ }\KeywordTok{italic}\NormalTok{(r)),}
  \DataTypeTok{ylab =} \KeywordTok{bquote}\NormalTok{(}\StringTok{"Density"} \OperatorTok{~}\StringTok{ }\KeywordTok{italic}\NormalTok{(f)),}
  \DataTypeTok{sub  =} \StringTok{"Pearle (black) vs. uniform in ball (green)"}\NormalTok{)}
\KeywordTok{lines}\NormalTok{(r, g, }\DataTypeTok{col =} \StringTok{"green"}\NormalTok{)}
\end{Highlighting}
\end{Shaded}

\includegraphics{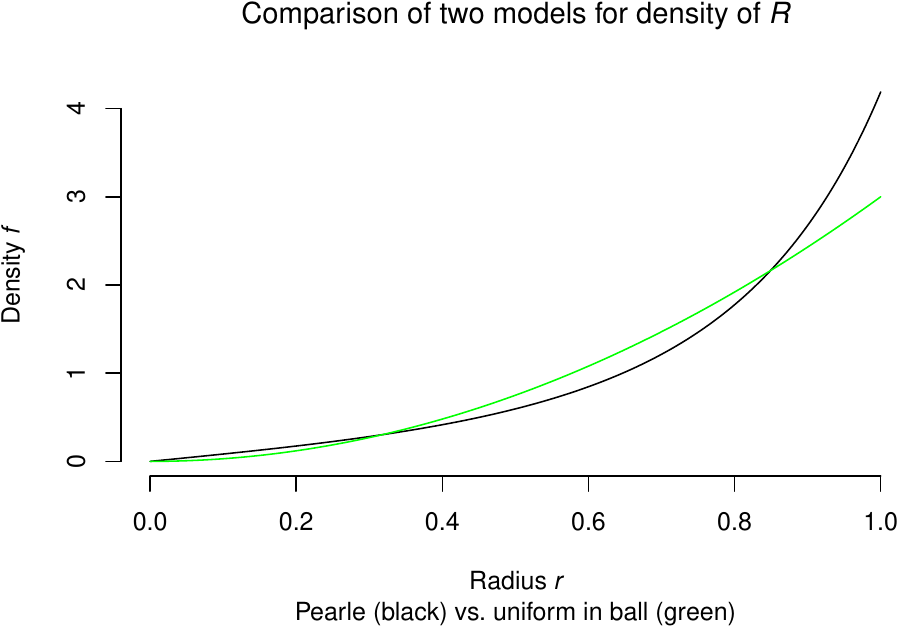}

We see that Pearle's points have a tendency to be closer to the surface
of the ball than if they had been uniformly distributed throughout it.

According to Pearle's model, Particle 1 is represented by a point in the
ball. It is \emph{observed} when its spin is measured in a certain
direction, if and only if its point lies in either of two ``mushroom
shaped'' regions around the measurement direction and its opposite. It
then delivers the \emph{outcome} +/-1 depending on the region. Particle
2 is represented by the exactly opposite point in the ball but for the
rest, its observation and measurement follow exactly the same rule. So
if both are measured in the same direction, either neither is observed,
or both are observed, and the two outcomes are opposite. Measured in
opposite directions, either neither will be observed, or both will be
observed, and the two outcomes will be the same.

In the following plot I draw the intersections of the two mushrooms (one
blue, with a plane through the origin containing the measurement
direction, which is taken to be the direction of the positive
\(x\)-axis. I superimpose on this plot a sample of \(1000\) particles
distributed in a circularly symmetric way about the centre of the unit
\emph{disk}, with distance to the origin distributed as in Pearle's
model. The result is a 2D \emph{caricature} of Pearle's 3D model: some
statistical features are the same, some are different.

The picture is neither a 2D section nor a 2D projection of the 3D model.
However, it should help the reader to visualise the model. The points
are coloured blue, red or black according to whether the corresponding
particle measurement result is an outcome spin up, spin down, or the
particle is not detected. Two particles are simultaneously measured in
this way: same point in the ball, different directions of measurement.

The actual detection regions (the red and the blue mushroom) are formed
by rotating the 2D boundaries about the \(x\)-axis. The actual
distribution of the 3D hidden variable of the particle being measured
has the same radial component as in the plot, but its direction is now
uniform over the sphere, instead of the circle. Thus the 3D density of
points is less than what the picture suggests as we move further from
the origin, however it still increases as we move outward relative to a
uniform density.

\begin{Shaded}
\begin{Highlighting}[]
\KeywordTok{set.seed}\NormalTok{(}\DecValTok{1234}\NormalTok{)}
\KeywordTok{par}\NormalTok{(}\DataTypeTok{pty =} \StringTok{"s"}\NormalTok{)}
\KeywordTok{plot}\NormalTok{(r }\OperatorTok{*}\StringTok{ }\KeywordTok{cos}\NormalTok{(r }\OperatorTok{*}\StringTok{ }\NormalTok{pi }\OperatorTok{/}\StringTok{ }\DecValTok{2}\NormalTok{), r }\OperatorTok{*}\StringTok{ }\KeywordTok{sin}\NormalTok{(r }\OperatorTok{*}\StringTok{ }\NormalTok{pi }\OperatorTok{/}\StringTok{ }\DecValTok{2}\NormalTok{), }\DataTypeTok{type =} \StringTok{"l"}\NormalTok{, }
  \DataTypeTok{main =} \StringTok{"Pearle's model, particle 1 measured in dir. pos. x-axis"}\NormalTok{,}
  \DataTypeTok{sub =} \StringTok{"Outcome +1 blue, -1 red, not observed black"}\NormalTok{,}
  \DataTypeTok{xlim =} \KeywordTok{c}\NormalTok{(}\OperatorTok{-}\DecValTok{1}\NormalTok{, }\DecValTok{1}\NormalTok{), }\DataTypeTok{ylim =} \KeywordTok{c}\NormalTok{(}\OperatorTok{-}\DecValTok{1}\NormalTok{, }\DecValTok{1}\NormalTok{), }\DataTypeTok{asp =} \DecValTok{1}\NormalTok{, }\DataTypeTok{col =} \StringTok{"blue"}\NormalTok{,}
  \DataTypeTok{xlab=}\StringTok{""}\NormalTok{,}
  \DataTypeTok{ylab=}\StringTok{""}\NormalTok{)}
\KeywordTok{lines}\NormalTok{(r }\OperatorTok{*}\StringTok{ }\KeywordTok{cos}\NormalTok{(r }\OperatorTok{*}\StringTok{ }\NormalTok{pi }\OperatorTok{/}\StringTok{ }\DecValTok{2}\NormalTok{), }\OperatorTok{-}\StringTok{ }\NormalTok{r }\OperatorTok{*}\StringTok{ }\KeywordTok{sin}\NormalTok{(r }\OperatorTok{*}\StringTok{ }\NormalTok{pi }\OperatorTok{/}\StringTok{ }\DecValTok{2}\NormalTok{), }\DataTypeTok{col =} \StringTok{"blue"}\NormalTok{)}
\KeywordTok{lines}\NormalTok{(}\OperatorTok{-}\StringTok{ }\NormalTok{r }\OperatorTok{*}\StringTok{ }\KeywordTok{cos}\NormalTok{(r }\OperatorTok{*}\StringTok{ }\NormalTok{pi }\OperatorTok{/}\StringTok{ }\DecValTok{2}\NormalTok{), r }\OperatorTok{*}\StringTok{ }\KeywordTok{sin}\NormalTok{(r }\OperatorTok{*}\StringTok{ }\NormalTok{pi }\OperatorTok{/}\StringTok{ }\DecValTok{2}\NormalTok{), }\DataTypeTok{col =} \StringTok{"red"}\NormalTok{)}
\KeywordTok{lines}\NormalTok{(}\OperatorTok{-}\StringTok{ }\NormalTok{r }\OperatorTok{*}\StringTok{ }\KeywordTok{cos}\NormalTok{(r }\OperatorTok{*}\StringTok{ }\NormalTok{pi }\OperatorTok{/}\StringTok{ }\DecValTok{2}\NormalTok{), }\OperatorTok{-}\StringTok{ }\NormalTok{r }\OperatorTok{*}\StringTok{ }\KeywordTok{sin}\NormalTok{(r }\OperatorTok{*}\StringTok{ }\NormalTok{pi }\OperatorTok{/}\StringTok{ }\DecValTok{2}\NormalTok{), }\DataTypeTok{col =} \StringTok{"red"}\NormalTok{)}
\NormalTok{t <-}\StringTok{ }\KeywordTok{seq}\NormalTok{(}\DataTypeTok{from =} \DecValTok{0}\NormalTok{, }\DataTypeTok{to =}\NormalTok{ pi, }\DataTypeTok{length =} \DecValTok{1000}\NormalTok{)}
\KeywordTok{lines}\NormalTok{(}\KeywordTok{sin}\NormalTok{(t), }\KeywordTok{cos}\NormalTok{(t), }\DataTypeTok{col =} \StringTok{"blue"}\NormalTok{)}
\KeywordTok{lines}\NormalTok{(}\OperatorTok{-}\StringTok{ }\KeywordTok{sin}\NormalTok{(t), }\KeywordTok{cos}\NormalTok{(t), }\DataTypeTok{col =} \StringTok{"red"}\NormalTok{)}
\NormalTok{S <-}\StringTok{ }\DecValTok{2} \OperatorTok{/}\StringTok{ }\KeywordTok{sqrt}\NormalTok{(}\DecValTok{1} \OperatorTok{+}\StringTok{ }\DecValTok{3} \OperatorTok{*}\StringTok{ }\KeywordTok{runif}\NormalTok{(}\DecValTok{1000}\NormalTok{)) }\OperatorTok{-}\StringTok{ }\DecValTok{1}
\NormalTok{R <-}\StringTok{ }\KeywordTok{acos}\NormalTok{(S) }\OperatorTok{/}\StringTok{ }\NormalTok{(pi}\OperatorTok{/}\DecValTok{2}\NormalTok{)}
\NormalTok{Theta <-}\StringTok{ }\KeywordTok{runif}\NormalTok{(}\DecValTok{1000}\NormalTok{) }\OperatorTok{*}\StringTok{ }\DecValTok{2} \OperatorTok{*}\StringTok{ }\NormalTok{pi}
\NormalTok{Sign <-}\StringTok{ }\NormalTok{(Theta }\OperatorTok{<}\StringTok{ }\NormalTok{pi}\OperatorTok{/}\DecValTok{2} \OperatorTok{|}\StringTok{ }\NormalTok{Theta }\OperatorTok{>}\StringTok{ }\DecValTok{3} \OperatorTok{*}\StringTok{ }\NormalTok{pi}\OperatorTok{/}\DecValTok{2}\NormalTok{)}
\NormalTok{Dev <-}\StringTok{ }\NormalTok{Theta }\OperatorTok{*}\StringTok{ }\NormalTok{(Theta }\OperatorTok{<}\StringTok{ }\NormalTok{pi}\OperatorTok{/}\DecValTok{2}\NormalTok{) }\OperatorTok{+}\StringTok{ }\NormalTok{(}\DecValTok{2} \OperatorTok{*}\StringTok{ }\NormalTok{pi }\OperatorTok{-}\StringTok{ }\NormalTok{Theta) }\OperatorTok{*}\StringTok{ }\NormalTok{(Theta }\OperatorTok{>}\StringTok{ }\DecValTok{3} \OperatorTok{*}\StringTok{ }\NormalTok{pi}\OperatorTok{/}\DecValTok{2}\NormalTok{) }\OperatorTok{+}
\StringTok{  }\KeywordTok{abs}\NormalTok{(Theta }\OperatorTok{-}\StringTok{ }\NormalTok{pi) }\OperatorTok{*}\StringTok{ }\NormalTok{(Theta }\OperatorTok{>=}\StringTok{ }\NormalTok{pi}\OperatorTok{/}\DecValTok{2} \OperatorTok{&}\StringTok{ }\NormalTok{Theta }\OperatorTok{<=}\StringTok{ }\DecValTok{3} \OperatorTok{*}\StringTok{ }\NormalTok{pi}\OperatorTok{/}\DecValTok{2}\NormalTok{)}
\NormalTok{Obs <-}\StringTok{ }\NormalTok{Dev }\OperatorTok{<}\StringTok{ }\NormalTok{R }\OperatorTok{*}\StringTok{ }\NormalTok{pi}\OperatorTok{/}\DecValTok{2}
\NormalTok{Col <-}\StringTok{ }\KeywordTok{rep}\NormalTok{(}\StringTok{"black"}\NormalTok{, }\DecValTok{1000}\NormalTok{)}
\NormalTok{Col[Obs }\OperatorTok{&}\StringTok{ }\NormalTok{Sign] <-}\StringTok{ "blue"}
\NormalTok{Col[Obs }\OperatorTok{&}\StringTok{ }\OperatorTok{!}\NormalTok{Sign] <-}\StringTok{ "red"}
\KeywordTok{points}\NormalTok{(R }\OperatorTok{*}\StringTok{ }\KeywordTok{cos}\NormalTok{(Theta), R }\OperatorTok{*}\StringTok{ }\KeywordTok{sin}\NormalTok{(Theta), }\DataTypeTok{pch =} \StringTok{"."}\NormalTok{, }\DataTypeTok{col =}\NormalTok{ Col)}
\end{Highlighting}
\end{Shaded}

\includegraphics{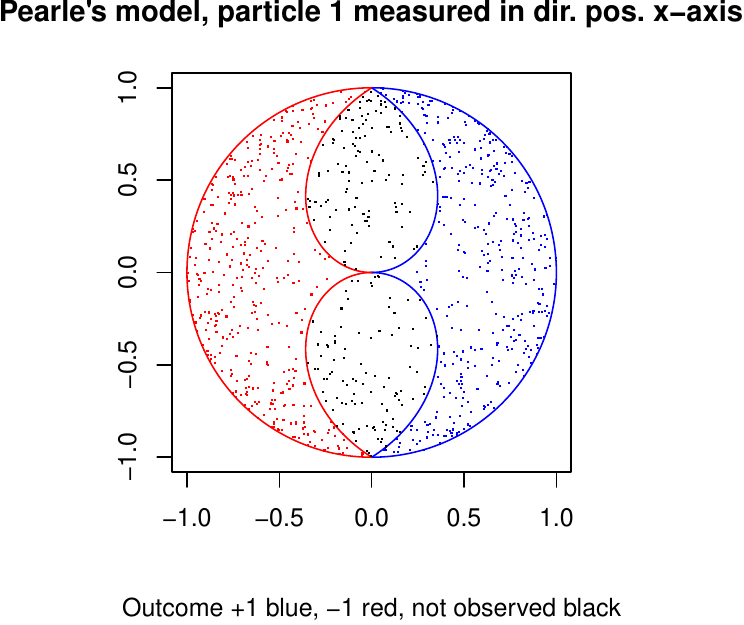}

Pearle found his model by fixing the mushroom shape first, then looking
for a probability distribution of the radial distance \(R\) such that
the pair (mushroom shape, distribution of \(R\)) reproduce the singlet
correlations. If we transform the unit ball onto itself in a continuous
way by applying a monotone transformation of the unit interval onto
itself to the distances of points from the origin, we can transform the
distribution of \(R\) into any distribution on \([0, 1]\) with
cumulative distribution function which is continuous and strictly
increasing throughout the interval. The mushroom shape will be
transformed correspondingly. The author of this paper has not discovered
a transformation which simultaneously makes both the the mushroom shape
and the distribution of \(R\) more simple than what they are at present.
In fact, Pearle's choice does amount to fixing one of these two coupled
parameters so that it has a direct physical interpretation: a particle
pair with a particular value \(r\) of \(R\) is such that each particle
is not detected at all if the direction of its spin, thought of now as
an undirected line through the origin, deviates by more than \(r\pi/2\)
from the direction in which the spin is measured, also thought of an
undirected line through the origin.

Altogether, Pearle's derivation of his model was a tour-de-force in
imagination, analysis and geometry. Whether or not there is a short-cut
to getting his results and whether or not they can be improved are
interesting challenges. As we will see in the next section, the model
has one major defect, namely the rate at which a pair of particles are
both detected depends quite strongly on the pair of settings with which
they are measured. This phenomenon would be experimentally observable;
conversely, the usual quantum mechanical modelling of this experiment,
and assuming that particles are detected independently of the direction
in which their spin is measured, predicts that the rate of pair
detection is independent of measurement settings. So we are left with
the open problem: is there a distribution of \(R\) reproducing the
singlet correlations which does not have this defect? Pearle does not
answer this question explicitly but his text suggests that he believes
the answer is negative. A numerical analysis (see Appendix) confirms.

\section{A simulation experiment}

We now present a simulation of the model in the statistical programming
language ``R''. First of all, we (re)set the random seed, for
reproducibility. To see results based on a fresh sample, replace the
(integer) seed by your own, or delete this line and let your computer
dream up one for you (it uses system time + process ID to do this job).

\begin{Shaded}
\begin{Highlighting}[]
\CommentTok{# set.seed() # Initialise random seed from system time + process ID}
\KeywordTok{set.seed}\NormalTok{(}\DecValTok{9875}\NormalTok{)  }\CommentTok{# Initialise random seed deterministically}
\end{Highlighting}
\end{Shaded}

We will generate uniform random points on sphere generated using the
``trig method'' (method 3) of Dave Seaman: see
\url{http://rpubs.com/gill1109/13340} for an R illustration. This very
effective but little known method uses the \emph{coincidence} that in
3D, a uniform point on the sphere has a \(z\) coordinate which is
uniformly distributed between \(-1\) and \(+1\). So we proceed as
follows.

\begin{enumerate}
\def\labelenumi{(\alph{enumi})}
\item
  Choose \(Z\) uniformly distributed in \([-1,1]\).
\item
  Choose \(\Theta\) uniformly distributed on \([0, 2\pi)\).
\item
  Let \(R = \sqrt(1-Z^2)\).
\item
  Let \(X = R \cos(\Theta)\).
\item
  Let \(Y = R \sin(\Theta)\).
\end{enumerate}

In the following simulation, the measurement directions will all be in
the equatorial plane, so only \(Z\) and \(X\) have been generated and
are treated as \(X\) and \(Y\).

First of all we set up the measurement angles for setting ``a'':
directions in the equatorial plane.

\begin{Shaded}
\begin{Highlighting}[]
\NormalTok{angles <-}\StringTok{ }\KeywordTok{seq}\NormalTok{(}\DataTypeTok{from =} \DecValTok{0}\NormalTok{, }\DataTypeTok{to =} \DecValTok{360}\NormalTok{, }\DataTypeTok{by =} \DecValTok{1}\NormalTok{) }\OperatorTok{*}\StringTok{ }\DecValTok{2} \OperatorTok{*}\StringTok{ }\NormalTok{pi}\OperatorTok{/}\DecValTok{360}
\NormalTok{K <-}\StringTok{ }\KeywordTok{length}\NormalTok{(angles)}
\NormalTok{corrs <-}\StringTok{ }\KeywordTok{numeric}\NormalTok{(K)  }\CommentTok{# Container for correlations}
\NormalTok{Ns <-}\StringTok{ }\KeywordTok{numeric}\NormalTok{(K)  }\CommentTok{# Container for number of states}
\end{Highlighting}
\end{Shaded}

For setting ``b'' we'll use a fixed direction.

\begin{Shaded}
\begin{Highlighting}[]
\NormalTok{beta <-}\StringTok{ }\DecValTok{0} \OperatorTok{*}\StringTok{ }\DecValTok{2} \OperatorTok{*}\StringTok{ }\NormalTok{pi}\OperatorTok{/}\DecValTok{360}  \CommentTok{# Measurement direction 'b' fixed, in equatorial plane}
\end{Highlighting}
\end{Shaded}

Then the sample size (number of pairs of particles).

\begin{Shaded}
\begin{Highlighting}[]
\NormalTok{M <-}\StringTok{ }\DecValTok{10}\OperatorTok{^}\DecValTok{6}
\end{Highlighting}
\end{Shaded}

I use the same, single sample of \(M = 10^6\) realizations of hidden
states for all measurement directions.

\begin{Shaded}
\begin{Highlighting}[]
\NormalTok{z <-}\StringTok{ }\KeywordTok{runif}\NormalTok{(M, }\DecValTok{-1}\NormalTok{, }\DecValTok{1}\NormalTok{)}
\NormalTok{t <-}\StringTok{ }\KeywordTok{runif}\NormalTok{(M, }\DecValTok{0}\NormalTok{, }\DecValTok{2} \OperatorTok{*}\StringTok{ }\NormalTok{pi)}
\NormalTok{r <-}\StringTok{ }\KeywordTok{sqrt}\NormalTok{(}\DecValTok{1} \OperatorTok{-}\StringTok{ }\NormalTok{z}\OperatorTok{^}\DecValTok{2}\NormalTok{)}
\NormalTok{x <-}\StringTok{ }\NormalTok{r }\OperatorTok{*}\StringTok{ }\KeywordTok{cos}\NormalTok{(t)}
\NormalTok{e <-}\StringTok{ }\KeywordTok{rbind}\NormalTok{(z, x)  }\CommentTok{# 2 x M matrix}
\end{Highlighting}
\end{Shaded}

The \(M\) columns of \(e\) represent the \(x\) and \(y\) coordinates of
\(M\) uniform random points on the sphere \(S^2\).

\begin{Shaded}
\begin{Highlighting}[]
\NormalTok{U <-}\StringTok{ }\KeywordTok{runif}\NormalTok{(M)}
\NormalTok{s <-}\StringTok{ }\NormalTok{(}\DecValTok{2}\OperatorTok{/}\KeywordTok{sqrt}\NormalTok{(}\DecValTok{3}\OperatorTok{*}\NormalTok{U}\OperatorTok{+}\DecValTok{1}\NormalTok{)) }\OperatorTok{-}\StringTok{ }\DecValTok{1}  \CommentTok{# Pearle's "r" is arc cosine of "s" divided by pi/2}
\NormalTok{b <-}\StringTok{ }\KeywordTok{c}\NormalTok{(}\KeywordTok{cos}\NormalTok{(beta), }\KeywordTok{sin}\NormalTok{(beta))  }\CommentTok{# Measurement vector 'b'}
\end{Highlighting}
\end{Shaded}

Loop through measurement vectors ``a'' (except last = 360 degrees =
first):

\begin{Shaded}
\begin{Highlighting}[]
\ControlFlowTok{for}\NormalTok{ (i }\ControlFlowTok{in} \DecValTok{1}\OperatorTok{:}\NormalTok{(K }\OperatorTok{-}\StringTok{ }\DecValTok{1}\NormalTok{)) \{}
\NormalTok{    alpha <-}\StringTok{ }\NormalTok{angles[i]}
\NormalTok{    a <-}\StringTok{ }\KeywordTok{c}\NormalTok{(}\KeywordTok{cos}\NormalTok{(alpha), }\KeywordTok{sin}\NormalTok{(alpha))  }\CommentTok{# Measurement vector 'a'}
\NormalTok{    ca <-}\StringTok{ }\KeywordTok{colSums}\NormalTok{(e }\OperatorTok{*}\StringTok{ }\NormalTok{a)  }\CommentTok{# Inner products of cols of 'e' with 'a'}
\NormalTok{    cb <-}\StringTok{ }\KeywordTok{colSums}\NormalTok{(e }\OperatorTok{*}\StringTok{ }\NormalTok{b)  }\CommentTok{# Inner products of cols of 'e' with 'b'}
\NormalTok{    good <-}\StringTok{ }\KeywordTok{abs}\NormalTok{(ca) }\OperatorTok{>}\StringTok{ }\NormalTok{s }\OperatorTok{&}\StringTok{ }\KeywordTok{abs}\NormalTok{(cb) }\OperatorTok{>}\StringTok{ }\NormalTok{s  }\CommentTok{# Select the 'states' }
\NormalTok{    N <-}\StringTok{ }\KeywordTok{sum}\NormalTok{(good)}
\NormalTok{    corrs[i] <-}\StringTok{ }\KeywordTok{sum}\NormalTok{(}\KeywordTok{sign}\NormalTok{(ca[good]) }\OperatorTok{*}\StringTok{ }\KeywordTok{sign}\NormalTok{(cb[good]))}\OperatorTok{/}\NormalTok{N}
\NormalTok{    Ns[i] <-}\StringTok{ }\NormalTok{N}
\NormalTok{\}}
\NormalTok{corrs[K] <-}\StringTok{ }\NormalTok{corrs[}\DecValTok{1}\NormalTok{]}
\NormalTok{Ns[K] <-}\StringTok{ }\NormalTok{Ns[}\DecValTok{1}\NormalTok{]}
\end{Highlighting}
\end{Shaded}

Now we are ready to make some plots of the results.

\begin{Shaded}
\begin{Highlighting}[]
\KeywordTok{plot}\NormalTok{(angles }\OperatorTok{*}\StringTok{ }\DecValTok{180}\OperatorTok{/}\NormalTok{pi, corrs, }\DataTypeTok{type =} \StringTok{"l"}\NormalTok{, }\DataTypeTok{col =} \StringTok{"blue"}\NormalTok{, }
     \DataTypeTok{main =} \StringTok{"Two correlation functions"}\NormalTok{, }
     \DataTypeTok{xlab =} \StringTok{"Angle (degrees)"}\NormalTok{, }\DataTypeTok{ylab =} \StringTok{"Correlation"}\NormalTok{)}
\KeywordTok{points}\NormalTok{(angles }\OperatorTok{*}\StringTok{ }\DecValTok{180}\OperatorTok{/}\NormalTok{pi, corrs, }\DataTypeTok{col =} \StringTok{"blue"}\NormalTok{, }\DataTypeTok{pch =} \StringTok{"."}\NormalTok{, }\DataTypeTok{cex =} \DecValTok{2}\NormalTok{)}
\KeywordTok{lines}\NormalTok{(angles }\OperatorTok{*}\StringTok{ }\DecValTok{180}\OperatorTok{/}\NormalTok{pi, }\KeywordTok{cos}\NormalTok{(angles), }\DataTypeTok{col =} \StringTok{"black"}\NormalTok{)}
\KeywordTok{points}\NormalTok{(angles }\OperatorTok{*}\StringTok{ }\DecValTok{180}\OperatorTok{/}\NormalTok{pi, }\KeywordTok{cos}\NormalTok{(angles), }\DataTypeTok{col =} \StringTok{"black"}\NormalTok{, }\DataTypeTok{pch =} \StringTok{"."}\NormalTok{, }\DataTypeTok{cex =} \DecValTok{2}\NormalTok{)}
\KeywordTok{legend}\NormalTok{(}\DataTypeTok{x =} \DecValTok{0}\NormalTok{, }\DataTypeTok{y =} \DecValTok{0}\NormalTok{, }\DataTypeTok{legend =} \KeywordTok{c}\NormalTok{(}\StringTok{"Pearle"}\NormalTok{, }\StringTok{"cosine"}\NormalTok{), }\DataTypeTok{text.col =} \KeywordTok{c}\NormalTok{(}\StringTok{"blue"}\NormalTok{, }
    \StringTok{"black"}\NormalTok{), }\DataTypeTok{lty =} \DecValTok{1}\NormalTok{, }\DataTypeTok{col =} \KeywordTok{c}\NormalTok{(}\StringTok{"blue"}\NormalTok{, }\StringTok{"black"}\NormalTok{))}
\end{Highlighting}
\end{Shaded}

\includegraphics{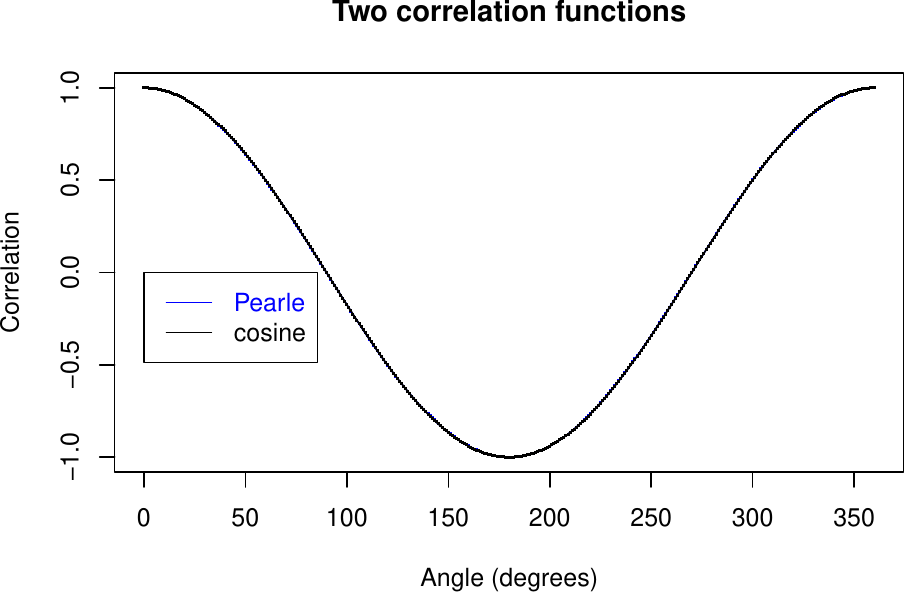}

In the second plot we zoom in on just part of the curve.

\begin{Shaded}
\begin{Highlighting}[]
\KeywordTok{plot}\NormalTok{(angles }\OperatorTok{*}\StringTok{ }\DecValTok{180}\OperatorTok{/}\NormalTok{pi, corrs, }\DataTypeTok{type =} \StringTok{"l"}\NormalTok{, }\DataTypeTok{col =} \StringTok{"blue"}\NormalTok{, }\DataTypeTok{xlim =} \KeywordTok{c}\NormalTok{(}\DecValTok{0}\NormalTok{, }\DecValTok{90}\NormalTok{), }\DataTypeTok{ylim =} \KeywordTok{c}\NormalTok{(}\DecValTok{0}\NormalTok{, }
    \DecValTok{1}\NormalTok{), }\DataTypeTok{main =} \StringTok{"Two correlation functions"}\NormalTok{, }\DataTypeTok{xlab =} \StringTok{"Angle (degrees)"}\NormalTok{, }\DataTypeTok{ylab =} \StringTok{"Correlation"}\NormalTok{)}
\KeywordTok{points}\NormalTok{(angles }\OperatorTok{*}\StringTok{ }\DecValTok{180}\OperatorTok{/}\NormalTok{pi, corrs, }\DataTypeTok{col =} \StringTok{"blue"}\NormalTok{, }\DataTypeTok{pch =} \StringTok{"."}\NormalTok{, }\DataTypeTok{cex =} \DecValTok{2}\NormalTok{)}
\KeywordTok{lines}\NormalTok{(angles }\OperatorTok{*}\StringTok{ }\DecValTok{180}\OperatorTok{/}\NormalTok{pi, }\KeywordTok{cos}\NormalTok{(angles), }\DataTypeTok{col =} \StringTok{"black"}\NormalTok{)}
\KeywordTok{points}\NormalTok{(angles }\OperatorTok{*}\StringTok{ }\DecValTok{180}\OperatorTok{/}\NormalTok{pi, }\KeywordTok{cos}\NormalTok{(angles), }\DataTypeTok{col =} \StringTok{"black"}\NormalTok{, }\DataTypeTok{pch =} \StringTok{"."}\NormalTok{, }\DataTypeTok{cex =} \DecValTok{2}\NormalTok{)}
\KeywordTok{legend}\NormalTok{(}\DataTypeTok{x =} \DecValTok{0}\NormalTok{, }\DataTypeTok{y =} \FloatTok{0.4}\NormalTok{, }\DataTypeTok{legend =} \KeywordTok{c}\NormalTok{(}\StringTok{"Pearle"}\NormalTok{, }\StringTok{"cosine"}\NormalTok{), }\DataTypeTok{text.col =} \KeywordTok{c}\NormalTok{(}\StringTok{"blue"}\NormalTok{, }
    \StringTok{"black"}\NormalTok{), }\DataTypeTok{lty =} \DecValTok{1}\NormalTok{, }\DataTypeTok{col =} \KeywordTok{c}\NormalTok{(}\StringTok{"blue"}\NormalTok{, }\StringTok{"black"}\NormalTok{))}
\end{Highlighting}
\end{Shaded}

\includegraphics{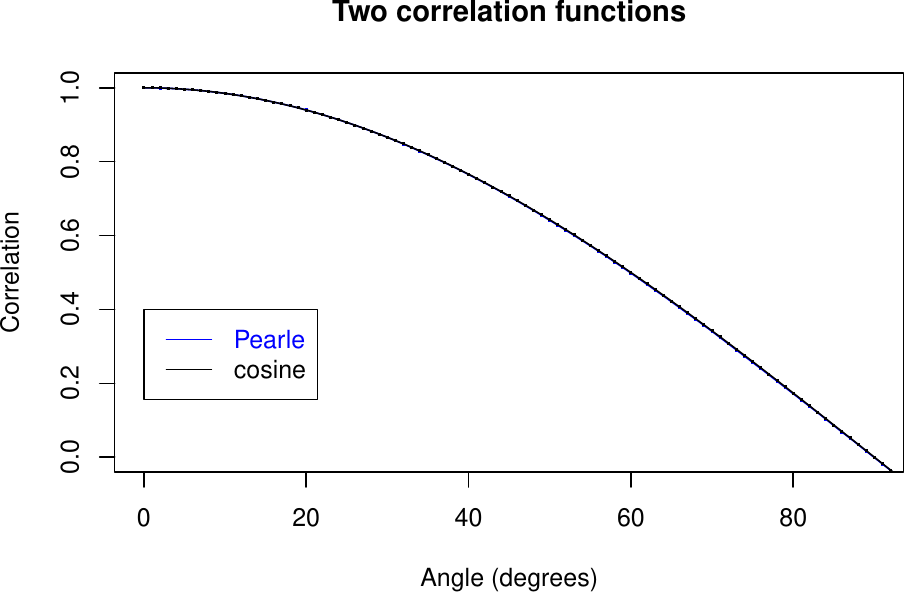}

Here is an even closer look at part of the curves.

\begin{Shaded}
\begin{Highlighting}[]
\KeywordTok{plot}\NormalTok{(angles }\OperatorTok{*}\StringTok{ }\DecValTok{180}\OperatorTok{/}\NormalTok{pi, corrs, }\DataTypeTok{type =} \StringTok{"l"}\NormalTok{, }\DataTypeTok{col =} \StringTok{"blue"}\NormalTok{, }
     \DataTypeTok{xlim =} \KeywordTok{c}\NormalTok{(}\DecValTok{0}\NormalTok{, }\DecValTok{50}\NormalTok{), }\DataTypeTok{ylim =} \KeywordTok{c}\NormalTok{(}\FloatTok{0.8}\NormalTok{, }\DecValTok{1}\NormalTok{), }
     \DataTypeTok{main =} \StringTok{"Two correlation functions"}\NormalTok{, }
     \DataTypeTok{xlab =} \StringTok{"Angle (degrees)"}\NormalTok{, }\DataTypeTok{ylab =} \StringTok{"Correlation"}\NormalTok{)}
\KeywordTok{points}\NormalTok{(angles }\OperatorTok{*}\StringTok{ }\DecValTok{180}\OperatorTok{/}\NormalTok{pi, corrs, }\DataTypeTok{type =} \StringTok{"b"}\NormalTok{, }\DataTypeTok{col =} \StringTok{"blue"}\NormalTok{, }\DataTypeTok{pch =} \StringTok{"."}\NormalTok{, }\DataTypeTok{cex =} \DecValTok{2}\NormalTok{)}
\KeywordTok{lines}\NormalTok{(angles }\OperatorTok{*}\StringTok{ }\DecValTok{180}\OperatorTok{/}\NormalTok{pi, }\KeywordTok{cos}\NormalTok{(angles), }\DataTypeTok{col =} \StringTok{"black"}\NormalTok{)}
\KeywordTok{points}\NormalTok{(angles }\OperatorTok{*}\StringTok{ }\DecValTok{180}\OperatorTok{/}\NormalTok{pi, }\KeywordTok{cos}\NormalTok{(angles), }\DataTypeTok{type =} \StringTok{"b"}\NormalTok{, }\DataTypeTok{col =} \StringTok{"black"}\NormalTok{, }\DataTypeTok{pch =} \StringTok{"."}\NormalTok{, }\DataTypeTok{cex =} \DecValTok{2}\NormalTok{)}
\KeywordTok{legend}\NormalTok{(}\DataTypeTok{x =} \DecValTok{0}\NormalTok{, }\DataTypeTok{y =} \FloatTok{0.85}\NormalTok{, }\DataTypeTok{legend =} \KeywordTok{c}\NormalTok{(}\StringTok{"Pearle"}\NormalTok{, }\StringTok{"cosine"}\NormalTok{), }\DataTypeTok{text.col =} \KeywordTok{c}\NormalTok{(}\StringTok{"blue"}\NormalTok{, }
    \StringTok{"black"}\NormalTok{), }\DataTypeTok{lty =} \DecValTok{1}\NormalTok{, }\DataTypeTok{col =} \KeywordTok{c}\NormalTok{(}\StringTok{"blue"}\NormalTok{, }\StringTok{"black"}\NormalTok{))}
\end{Highlighting}
\end{Shaded}

\includegraphics{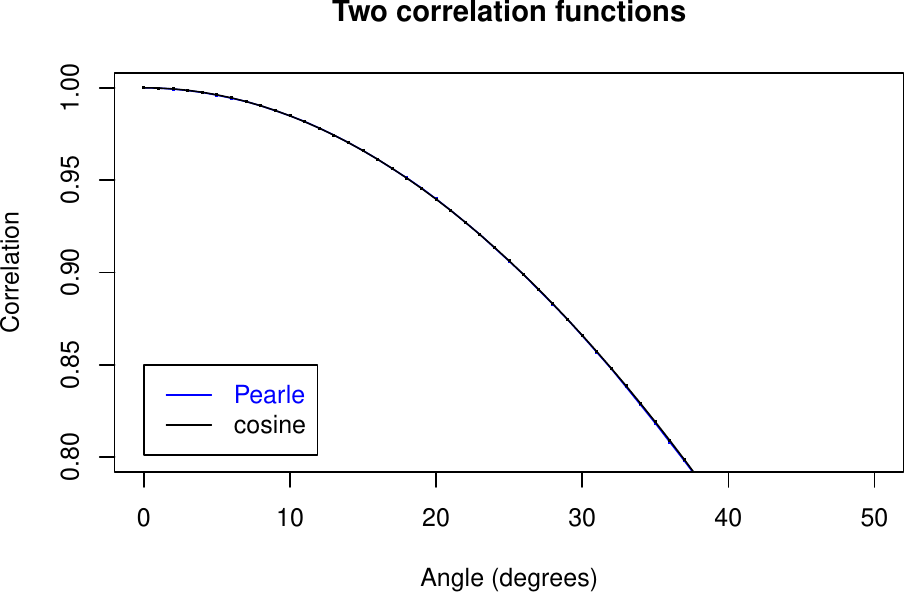}

Here is a plot of the differences between theory and simulation, with an
indication of accuracy.

\begin{Shaded}
\begin{Highlighting}[]
\KeywordTok{plot}\NormalTok{(angles }\OperatorTok{*}\StringTok{ }\DecValTok{180}\OperatorTok{/}\NormalTok{pi, corrs }\OperatorTok{-}\StringTok{ }\KeywordTok{cos}\NormalTok{(angles), }\DataTypeTok{type =} \StringTok{"l"}\NormalTok{, }\DataTypeTok{col =} \StringTok{"blue"}\NormalTok{,}
     \DataTypeTok{main =} \StringTok{"Difference (red: upper bound to +/- 1 standard error)"}\NormalTok{)}
\KeywordTok{abline}\NormalTok{(}\DataTypeTok{h =} \DecValTok{0}\NormalTok{, }\DataTypeTok{col =} \StringTok{"black"}\NormalTok{, }\DataTypeTok{lwd =} \DecValTok{2}\NormalTok{)}
\KeywordTok{lines}\NormalTok{(angles }\OperatorTok{*}\StringTok{ }\DecValTok{180}\OperatorTok{/}\NormalTok{pi, }\DecValTok{1}\OperatorTok{/}\KeywordTok{sqrt}\NormalTok{(Ns), }\DataTypeTok{col =} \StringTok{"red"}\NormalTok{)}
\KeywordTok{lines}\NormalTok{(angles }\OperatorTok{*}\StringTok{ }\DecValTok{180}\OperatorTok{/}\NormalTok{pi, }\DecValTok{-1}\OperatorTok{/}\KeywordTok{sqrt}\NormalTok{(Ns), }\DataTypeTok{col =} \StringTok{"red"}\NormalTok{)}
\end{Highlighting}
\end{Shaded}

\includegraphics{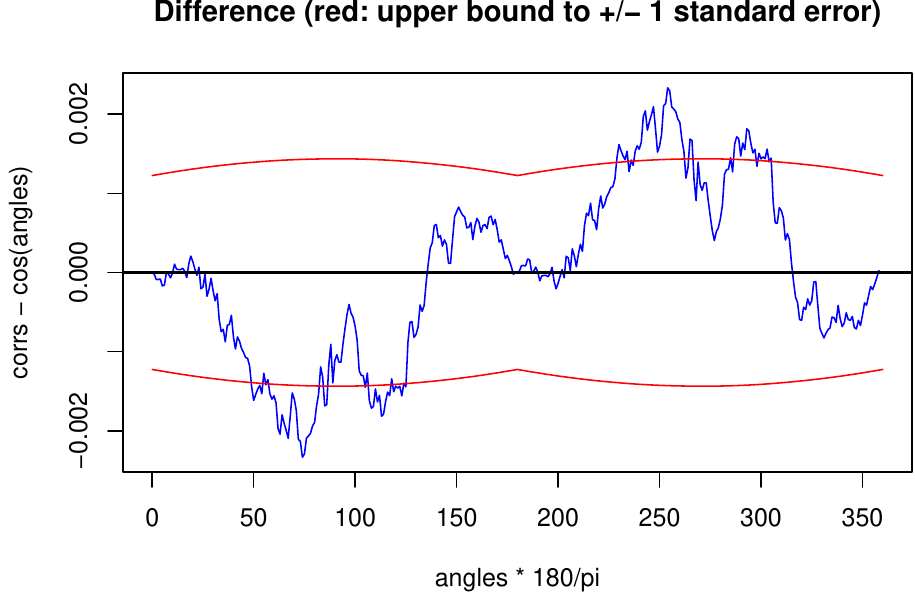}

\begin{Shaded}
\begin{Highlighting}[]
\KeywordTok{max}\NormalTok{(}\KeywordTok{abs}\NormalTok{(corrs }\OperatorTok{-}\StringTok{ }\KeywordTok{cos}\NormalTok{(angles)))}
\end{Highlighting}
\end{Shaded}

\begin{verbatim}
## [1] 0.002333192
\end{verbatim}

Finally, a plot of the proportion of observed particle pairs to emitted
pairs, as function of the angle between the measurement directions.

\begin{Shaded}
\begin{Highlighting}[]
\KeywordTok{plot}\NormalTok{(angles }\OperatorTok{*}\StringTok{ }\DecValTok{180}\OperatorTok{/}\NormalTok{pi, Ns }\OperatorTok{/}\StringTok{ }\NormalTok{M, }\DataTypeTok{type =} \StringTok{"l"}\NormalTok{, }\DataTypeTok{col =} \StringTok{"blue"}\NormalTok{,}
     \DataTypeTok{main =} \StringTok{"Rate of detected particle pairs"}\NormalTok{, }\DataTypeTok{ylim =} \KeywordTok{c}\NormalTok{(}\DecValTok{0}\NormalTok{, }\DecValTok{1}\NormalTok{))}
\KeywordTok{abline}\NormalTok{(}\DataTypeTok{h =}\NormalTok{ (}\DecValTok{2}\OperatorTok{/}\DecValTok{3}\NormalTok{))}
\KeywordTok{abline}\NormalTok{(}\DataTypeTok{h =}\NormalTok{ (}\DecValTok{4}\OperatorTok{/}\DecValTok{3}\NormalTok{) }\OperatorTok{*}\StringTok{ }\NormalTok{(}\DecValTok{1} \OperatorTok{-}\StringTok{ }\DecValTok{2}\OperatorTok{/}\NormalTok{pi))}
\end{Highlighting}
\end{Shaded}

\includegraphics{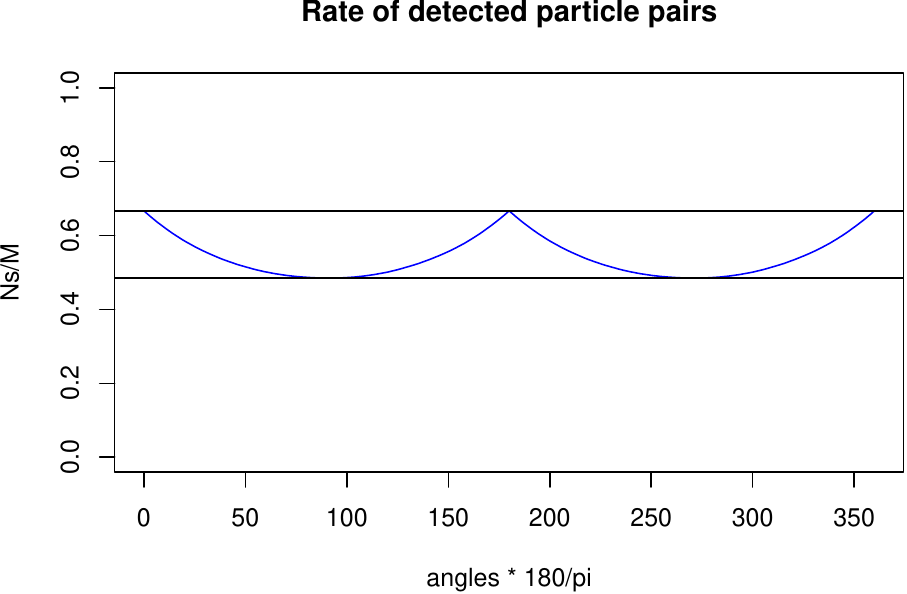}

The two horizontal lines are the maximum and minimum detection rates
computed by Pearle: \(2/3\) and \(4/3 (1 - 2/\pi) = 0.4845\dots\) of
\(M\) respectively. Now, if an experimenter is not using pulsed emission
of particle pairs but they are being emitted in a continuous fashion
according to a Poisson process, then the experimenter will have no way
of knowing that when neither particle is detected, there was still an
emission of a particle pair. So the loss of \(1/3\) of all emitted
particle pairs will go unnoticed. But the experimenter will be able to
see that the rate of double detections depends strongly on the
difference between the two measurement directions -- the maximum rate is
more than \(4/3\) times the smallest. Put another way: the rate at which
particles are detected at one measurement station with no accompanying
detection at the other depends on the difference between the two
measurement directions. Thus Pearle's model has some very unsatisfactory
features: assuming a constant emission rate, the experimenter can see
that particles are suspiciously being rejected in a way which depends on
both the settings. It was only in 2008 that Gisin and Gisin came up with
a new local hidden variable model for the singlet correlations based on
data rejection which possesses \emph{all} the symmetries one would
require. Moreover, it is amazingly simple. However it seems further from
physical interpretation than Pearle's model.

But Pearle did more than exhibit one concrete local hidden variable
model which reproduces the singlet correlations: he also characterizes
the class of all distributions of \(R\) which does the job. This allows
us in principle to find out if there is a distribution within the class
which leads to a model with all required symmetries. I believe the
answer is negative (and I believe that Pearle knew this too) but I do
not have a mathematical proof. Numerical evidence (see Appendix) is very
strong and inspection of the numerical result might help in constructing
a proof.

\section{Acknowledgements}

I was stimulated to figure out exactly what Pearle (1970) had done
during discussion on internet fora with Michel Fodje, Chantal Roth, Joy
Christian, and others. Michel Fodje had come up with his own detection
loophole simulation model and I started by comparing this with the
similar ``chaotic ball'' model of Caroline Thompson, see the arXiv preprint
Thompson and Holstein (2002). Thompson wrote a whole series of papers
on this topic but only ever got one paper published, Thompson (1996). 
She discusses Pearle's model at length. At least her work
is preserved on arXiv. In an interesting survey, Risco-Delgado (1993) 
also gives it a lot of attention, and includes a very nice 
picture explaining the idea of the model. 
His text simply copies Pearle's incorrect formulas.
My own versions of all these models, programmed in R, can
be found at my RPubs website \url{http://rpubs.com/gill1109}.

Florin Moldoveanu helped check my decoding of Pearle's derivation of the
density of \(R\). There is some ambiguity of notation (Pearle's notion
of ``probability density'' is unconventional by modern standards and
moreover seems not entirely consistent throughout the paper). This is
probably how the normalization error in the final result crept in,
midway through the computations. Since the error does not seem to have
been reported elsewhere, and since it becomes manifest as soon as one
attempts to implement a simulation of the model, I believe that this was
the first time anyone did actually attempt to simulate the Pearle model.
The simulation reported here was posted to RPubs in early March 2014.

\section*{Appendix}

Here I reproduce Pearle's description of the class of \emph{all}
distributions of \(R\) such that the singlet correlation is recovered
from the measurement outcomes of detected particle pairs.

Let \(\mu\) be a real function on \([0, 1]\) satisfying the symmetry
requirement \(\mu(x) = \mu((1 - x^2)^{1/2})\) for all \(x\). One could
for instance pick \(\mu\) arbitrarily on \([0, 1/\sqrt 2]\) and use the
symmetry requirement to determine \(\mu\) on \((1/\sqrt 2, 1]\). It
would probably be wise to impose continuity of the derivative of \(\mu\)
at \(1/\sqrt 2\). Next compute the function \(h\) through
\[h(x) ~=~ \frac {\textrm d ^2} {\textrm d x^2} 
\Biggl(\frac{x^2}{1 - x^2}\Biggl[\int_0^1 (1 - z^2)^{\frac12}\mu(z)\textrm d z - \int_0^x\Bigl(1 - \frac {z^2}{x^2}\Bigr)^{\frac12}\mu(z)\textrm d z\Biggr]\Biggr).\]
If \(h\) is nonnegative and integrable, normalize it to a probability
density: this should be the probability density of \(S = \cos(R\pi/2)\).
The choice \(\mu =\) constant delivers the particular distribution of
\(S\) which Pearle further investigates and which we have studied here.

I am not aware of a simple interpretation of the function \(\mu\) so its
role is hard to understand. It is the result of applying a certain
differential operator, Pearle's equation (17) \[
\mu(x)~=~-\frac 1 {(1 - x^2)^{1/2}}\frac{\mathrm d}{\mathrm d x}
\Biggl[\frac{(1 - x^2)^2}{4 x}
\frac{\mathrm d}{\mathrm d x}\Biggr(
g(x) x^2
\Biggr)
\Biggr]
\] to the function \(g\) defined as the probability of detection of a
particle pair, expressed as function of \(s = \cos(r\pi/2)\).

In principle we can therefore see what happens if we specify \(g =\)
constant, and giving \(\mu(x) = C x(1 - x^2)^{1/2}\); this results in a
candidate for the function \(h\) and we only have to find out whether or
not \(h\) can be normalized to a probability density (integrable,
nonnegative). I was not able to perform this operation analytically.
However, numerical integration and differentiation delivers us a
candidate \(h\) which takes both negative and positive values. This
provides strong evidence that his model does \emph{not} include a
distribution for \(R\) (or equivalently \(S\)) such that the pair
detection probability is independent of the settings. The numerical
analysis did confirm the theoretical analysis for the case \(\mu=\)
constant, so the author does have some faith in its results.

The investigation is hampered by the misprints in Pearle's paper:
for instance the power in equation (21) should be \(-3\) not \(+3\), and
throughout, normalisation constants are not to be trusted. The following
very naive code ``computes'' the density of \(S\) first in the case of
constant \(\mu\) then in the case of constant \(g\).

\begin{Shaded}
\begin{Highlighting}[]
\KeywordTok{rm}\NormalTok{(}\DataTypeTok{list =} \KeywordTok{ls}\NormalTok{()) }\CommentTok{# clear workspace}
\NormalTok{eps <-}\StringTok{ }\DecValTok{10}\OperatorTok{^}\NormalTok{\{}\OperatorTok{-}\DecValTok{9}\NormalTok{\}}
\NormalTok{Npts <-}\StringTok{ }\DecValTok{10}\OperatorTok{^}\DecValTok{4}
\NormalTok{z <-}\StringTok{ }\KeywordTok{seq}\NormalTok{(}\DataTypeTok{from =}\NormalTok{ eps, }\DataTypeTok{to =} \DecValTok{1} \OperatorTok{-}\StringTok{ }\NormalTok{eps, }\DataTypeTok{length =}\NormalTok{ Npts)}
\NormalTok{x <-}\StringTok{ }\NormalTok{z}

\NormalTok{Kernel0 <-}\StringTok{ }\KeywordTok{outer}\NormalTok{(z}\OperatorTok{^}\DecValTok{2}\NormalTok{, x}\OperatorTok{^}\DecValTok{2}\NormalTok{, }\StringTok{"/"}\NormalTok{)}
\NormalTok{Kernel0[}\KeywordTok{lower.tri}\NormalTok{(Kernel0)] <-}\StringTok{ }\DecValTok{1}
\NormalTok{Kernel0 <-}\StringTok{ }\KeywordTok{sqrt}\NormalTok{(}\DecValTok{1} \OperatorTok{-}\StringTok{ }\NormalTok{Kernel0)}

\NormalTok{mu <-}\StringTok{ }\DecValTok{1}

\NormalTok{First <-}\StringTok{ }\KeywordTok{mean}\NormalTok{(}\KeywordTok{sqrt}\NormalTok{(}\DecValTok{1} \OperatorTok{-}\StringTok{ }\NormalTok{z}\OperatorTok{^}\DecValTok{2}\NormalTok{) }\OperatorTok{*}\StringTok{ }\NormalTok{mu)}
\NormalTok{Kernel <-}\StringTok{ }\NormalTok{Kernel0 }\OperatorTok{*}\StringTok{ }\NormalTok{mu}
\NormalTok{Second <-}\StringTok{ }\KeywordTok{colMeans}\NormalTok{(Kernel)}
\NormalTok{result <-}\StringTok{ }\NormalTok{x}\OperatorTok{^}\DecValTok{2} \OperatorTok{*}\StringTok{ }\NormalTok{(First }\OperatorTok{-}\StringTok{ }\NormalTok{Second) }\OperatorTok{/}\StringTok{ }\NormalTok{(}\DecValTok{1} \OperatorTok{-}\StringTok{ }\NormalTok{x}\OperatorTok{^}\DecValTok{2}\NormalTok{)}

\NormalTok{dens <-}\StringTok{ }\KeywordTok{diff}\NormalTok{(}\KeywordTok{diff}\NormalTok{(result[}\DecValTok{1}\OperatorTok{:}\NormalTok{(Npts }\OperatorTok{-}\StringTok{ }\DecValTok{100} \OperatorTok{+}\StringTok{ }\DecValTok{2}\NormalTok{)]) }\OperatorTok{*}\StringTok{ }\NormalTok{Npts) }\OperatorTok{*}\StringTok{ }\NormalTok{Npts}
\NormalTok{s <-}\StringTok{ }\NormalTok{x[}\DecValTok{1} \OperatorTok{:}\StringTok{ }\NormalTok{(Npts }\OperatorTok{-}\StringTok{ }\DecValTok{100}\NormalTok{)] }
\CommentTok{# "-100" because of numerical instability at right endpoint}
\NormalTok{dens <-}\StringTok{ }\NormalTok{dens}\OperatorTok{/}\KeywordTok{mean}\NormalTok{(dens)}
\KeywordTok{plot}\NormalTok{(s, dens, }\DataTypeTok{type =} \StringTok{"l"}\NormalTok{, }\DataTypeTok{ylim =} \KeywordTok{c}\NormalTok{(}\DecValTok{0}\NormalTok{, }\KeywordTok{max}\NormalTok{(dens)),}
     \DataTypeTok{main =} \StringTok{"Solving for density numerically, with mu = constant"}\NormalTok{,}
     \DataTypeTok{xlab =} \StringTok{"radius"}\NormalTok{,}
     \DataTypeTok{ylab =} \StringTok{"density"}\NormalTok{,}
     \DataTypeTok{sub =} \StringTok{"In green, Pearle's analytical solution"}\NormalTok{)}
\CommentTok{# Note vertical offset 0.02 to separate curves:}
\KeywordTok{lines}\NormalTok{(s, }\FloatTok{0.02} \OperatorTok{+}\StringTok{ }\NormalTok{(}\DecValTok{1} \OperatorTok{+}\StringTok{ }\NormalTok{s)}\OperatorTok{^}\NormalTok{(}\OperatorTok{-}\DecValTok{3}\NormalTok{) }\OperatorTok{/}\StringTok{ }\KeywordTok{mean}\NormalTok{( (}\DecValTok{1} \OperatorTok{+}\StringTok{ }\NormalTok{s)}\OperatorTok{^}\NormalTok{(}\OperatorTok{-}\DecValTok{3}\NormalTok{) ), }\DataTypeTok{col =} \StringTok{"green"}\NormalTok{)}
\end{Highlighting}
\end{Shaded}

\includegraphics{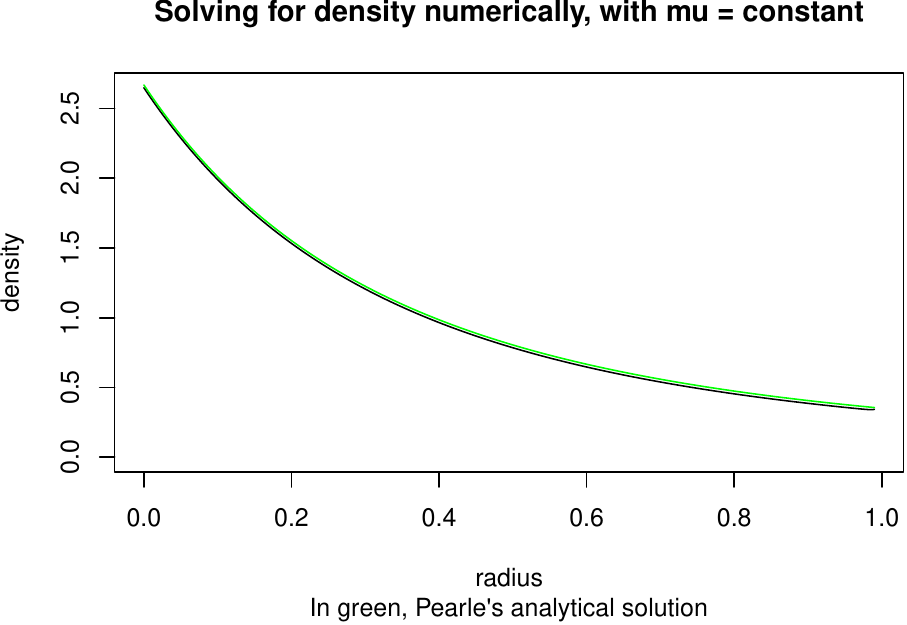}

\begin{Shaded}
\begin{Highlighting}[]
\NormalTok{mu <-}\StringTok{ }\NormalTok{z }\OperatorTok{*}\StringTok{ }\KeywordTok{sqrt}\NormalTok{(}\DecValTok{1} \OperatorTok{-}\StringTok{ }\NormalTok{z}\OperatorTok{^}\DecValTok{2}\NormalTok{)}

\NormalTok{First <-}\StringTok{ }\KeywordTok{mean}\NormalTok{(}\KeywordTok{sqrt}\NormalTok{(}\DecValTok{1} \OperatorTok{-}\StringTok{ }\NormalTok{z}\OperatorTok{^}\DecValTok{2}\NormalTok{) }\OperatorTok{*}\StringTok{ }\NormalTok{mu)}
\NormalTok{Kernel <-}\StringTok{ }\NormalTok{Kernel0 }\OperatorTok{*}\StringTok{ }\NormalTok{mu}
\NormalTok{Second <-}\StringTok{ }\KeywordTok{colMeans}\NormalTok{(Kernel)}
\NormalTok{result <-}\StringTok{ }\NormalTok{x}\OperatorTok{^}\DecValTok{2} \OperatorTok{*}\StringTok{ }\NormalTok{(First }\OperatorTok{-}\StringTok{ }\NormalTok{Second) }\OperatorTok{/}\StringTok{ }\NormalTok{(}\DecValTok{1} \OperatorTok{-}\StringTok{ }\NormalTok{x}\OperatorTok{^}\DecValTok{2}\NormalTok{)}

\NormalTok{dens <-}\StringTok{ }\KeywordTok{diff}\NormalTok{(}\KeywordTok{diff}\NormalTok{(result[}\DecValTok{1}\OperatorTok{:}\NormalTok{(Npts }\OperatorTok{-}\StringTok{ }\DecValTok{100} \OperatorTok{+}\StringTok{ }\DecValTok{2}\NormalTok{)]) }\OperatorTok{*}\StringTok{ }\NormalTok{Npts) }\OperatorTok{*}\StringTok{ }\NormalTok{Npts}
\NormalTok{s <-}\StringTok{ }\NormalTok{x[}\DecValTok{1} \OperatorTok{:}\StringTok{ }\NormalTok{(Npts }\OperatorTok{-}\StringTok{ }\DecValTok{100}\NormalTok{)]}
\NormalTok{dens <-}\StringTok{ }\NormalTok{dens}\OperatorTok{/}\KeywordTok{mean}\NormalTok{(dens)}
\KeywordTok{plot}\NormalTok{(s, dens, }\DataTypeTok{type =} \StringTok{"l"}\NormalTok{,}
     \DataTypeTok{main =} \StringTok{"Solving for density numerically, with g = constant"}\NormalTok{,}
     \DataTypeTok{xlab =} \StringTok{"radius"}\NormalTok{,}
     \DataTypeTok{ylab =} \StringTok{"density"}\NormalTok{,}
     \DataTypeTok{sub=}\StringTok{"Note that 'density' becomes negative"}\NormalTok{)}
\KeywordTok{abline}\NormalTok{(}\DataTypeTok{h =} \DecValTok{0}\NormalTok{)}
\end{Highlighting}
\end{Shaded}

\includegraphics{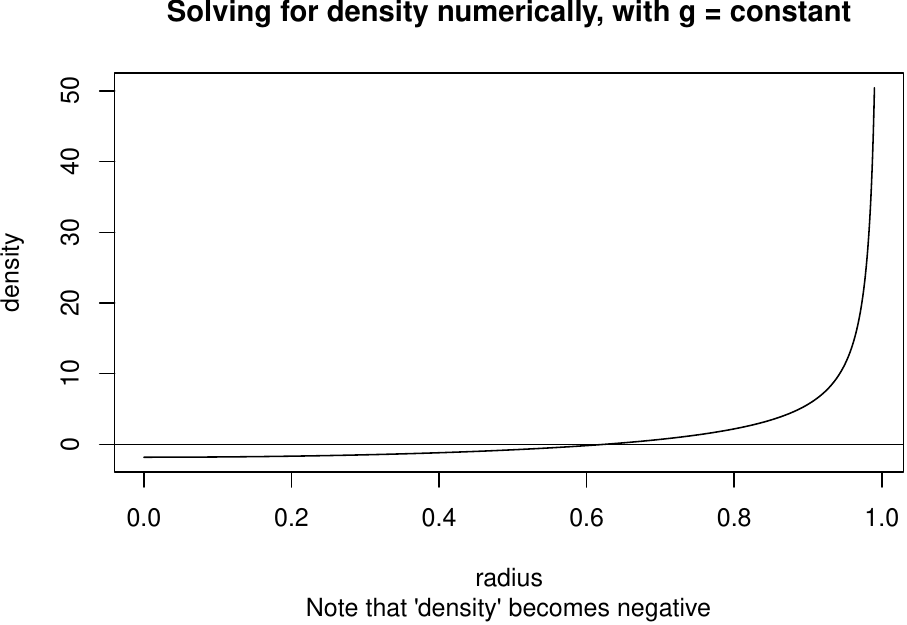}

\section*{References}

\raggedright
\frenchspacing

J.S. Bell (1964), On the {E}instein {P}odolsky {R}osen paradox.
\emph{Physics} \textbf{1} (3), 195--200. 

J.F. Clauser and M.A. Horne (1974), Experimental consequences of objective local theories.
\emph{Phys. Rev. D} \textbf{10} (2), 526--35.

J.F. Clauser, M.A. Horne, A. Shimony and R.A. Holt (1969), Proposed experiment to test local hidden-variable theories. \emph{Phys. Rev. Lett.} \textbf{23} (15), 880--4.

A. Garg and N.D. Mermin (1987), Detector inefficiencies in the {E}instein-{P}odolsky-{R}osen experiment. 
\emph{Physical Review D} \textbf{35} (12) 3831--5.

N. Gisin and B. Gisin (1999), A local hidden variable model of quantum
correlation exploiting the detection loophole. \emph{Phys. Lett. A}
\textbf{260}, 323--327.

J.-\AA. Larsson (1998), Bell's inequality and detector inefficiency. 
\emph{Phys. Rev. A} \textbf{57} 3304--8.

J.-\AA. Larsson (2014), Loopholes in {B}ell inequality tests of local realism.
\emph{J. Phys. A: Math. Theor.} \textbf{47} 424003.

J.-\AA. Larsson and J. Semitecolos (2001). Strict detector-efficiency bounds for $n$-site Clauser–Horne inequalities. 
\emph{Phys. Rev. A} \textbf{63} 022117.

P. Pearle (1970), Hidden-variable example based upon data rejection.
\emph{Phys. Rev.~D} \textbf{2}, 1418--1425.

R. Risto-Delgado (1993), The variable detection approach: a wave particle model.
\emph{Found. Phys. Lett.} \textbf{6} 399--428.

C.H. Thompson (1996) The chaotic ball: an intuitive analogy for EPR experiments.
\emph{Foundations of Physics Letters} \textbf{9} (4) 357--382.

C.H. Thompson and H. Holstein (2002), The ``Chaotic Ball'' model, local realism and the Bell test loopholes.
\url{http://arxiv.org/abs/quant-ph/0210150}.

\end{document}